\documentclass[useAMS,usenatbib]{mn2e} 
\usepackage{graphics}
\usepackage{epsfig}  
\usepackage{natbib} 
\usepackage{float}
\newcommand{\hMpc}{{\ifmmode{h^{-1}{\rm Mpc}}\else{$h^{-1}$Mpc }\fi}}
\newcommand{\hkpc}{{\ifmmode{h^{-1}{\rm kpc}}\else{$h^{-1}$kpc }\fi}}
\newcommand{\hMsun}{{\ifmmode{h^{-1}{\rm {M_{\odot}}}}\else{$h^{-1}{\rm{M_{\odot}}}$}\fi}}

\newcommand{\mgap}{\ensuremath{\Delta m_{12}}}

\title[Fossil Groups]{The Fossil Phase in the Life of a Galaxy Group} 
\author[von Benda-Beckmann et al.]{ Alexander M. von
Benda-Beckmann$^{1}$, Elena D'Onghia$^{2}$\thanks{Marie Curie
Fellow}, Stefan Gottl{\" o}ber$^{1}$, \newauthor Matthias
Hoeft$^{3}$, Arman Khalatyan$^{1}$, Anatoly Klypin$^{4}$, 
and Volker M{\" u}ller$^{1}$
\\
$^{1}$Astrophysical Institute Potsdam, An der Sternwarte 16, Germany \\
$^{2}$Institute for Theoretical Physics - University of Zurich,
Winterthurerstrasse 190, Switzerland\\
$^{3}$Jacobs University Bremen, Campus Ring 12, Germany \\
$^{4}$ Astronomy Department, New Mexico State University,
MSC 4500, P.O.Box 30001, Las Cruces, NM, 880003-8001, USA}

\begin{document}

\date{Submitted Version 2007 October}

\pagerange{\pageref{firstpage}--\pageref{lastpage}} \pubyear{2007}

\maketitle

\label{firstpage}

\begin{abstract}
 
  We investigate the origin and evolution of fossil groups in a concordance
  LCDM cosmological simulation. We consider haloes with masses between
  $(1-5)\times10^{13} \hMsun$ and study the physical mechanisms that lead to
  the formation of the large gap in magnitude between the brightest and the
  second most bright group member, which is typical for these fossil systems.
  Fossil groups are found to have high dark matter concentrations, which we
  can relate to their early formation time.  The large magnitude-gaps arise
  after the groups have build up half of their final mass, due to merging of
  massive group members. We show that the existence of fossil systems is
  primarily driven by the relatively early infall of massive satellites, and
  that we do not find a strong environmental dependence for these systems. In
  addition, we find tentative evidence for fossil group satellites falling in
  on orbits with typically lower angular momentum, which might lead to a more
  efficient merger onto the host. We find a population of groups at higher
  redshifts that go through a ``fossil phase'': a stage where they show a
  large magnitude-gap, which is terminated by renewed infall from their
  environment.

\end{abstract}


\section{INTRODUCTION}

Observations in the last decade revealed the existence of groups of galaxies
containing extended X-ray-emitting hot gas with properties expected for poor
clusters such as the Virgo cluster, but in the optical light completely
dominated by a single luminous, giant elliptical galaxy \cite[][]{Ponman1994a,
  Vikhlinin1999a}. The second brightest galaxy in these systems is more than a
factor five less luminous than the dominant elliptical.

To be specific, these systems are defined as spatially extended X-ray sources
with luminosities $L_{\rm X,bol} \ge 10^{42} h_{50}\;{\rm erg}\; {\rm s}^{-1}$.
The optical counterparts are galaxy groups with $\Delta m_{12} \ge 2 $ mag,
where $\Delta m_{12}$ is the absolute R-band magnitude-gap between the
brightest and second-brightest galaxies.

These systems are extremely interesting for several reasons: Although they
have X-ray temperatures comparable to the Virgo cluster these systems show a
galaxy luminosity function with a deficit of bright galaxies beyond the
characteristic magnitude of the Schechter function M* , or of visible galaxies
as compared to the predictions of cosmological simulations
\cite[][]{DOnghia2004a}, whereas Virgo contains six M* galaxies
\cite[][]{Jones2000a}.  Therefore they have been interpreted as the final
outcome of galaxy-galaxy mergers.  Numerical simulations suggest that the
luminous galaxies in a group will eventually merge to form a single giant
elliptical galaxy \cite[e.g.][]{Barnes1989a}. The merging timescales for the
brightest group members (with magnitudes M $\sim$ M* or brighter) in compact
groups are typically a few tenths of a Hubble time.  Therefore, by the present
day several group galaxies have likely merged into the giant elliptical.
Outside of the high-density core, the cooling time for the intra-group medium
is larger than a Hubble time; thus, while the luminous galaxies in some groups
have had enough time to merge into a single object, the large-scale X-ray halo
of the original groups should remain intact. This means that a merged group
might appear today as an isolated elliptical galaxy with a group-like X-ray
halo \cite[][]{Ponman1993a}. Hence these systems have been termed ``fossil''
groups.
  
Using the ROSAT All-Sky Survey data, \citet[][]{Ponman1994a} found the first
``fossil'' group candidate. The RXJ1340.6+4018 system has an X-ray luminosity
comparable to a group, but 70\% of the optical light comes from a single
elliptical galaxy \cite[][]{Jones2000a}. The galaxy luminosity function of
RXJ1340.6+4018 indicates a deficit of galaxies at the characteristic magnitude
M*. \citet[][]{Jones2000a}  have studied the central galaxy in detail and
found no evidence for spectral features implying recent star formation, which
indicates the last major merger occurred at least several Gyrs ago.
  
These systems are not rare. With a number density of $(5 \times 10^{-7}$ -- $2
\times 10^{-6}) \; h^{3} {\rm{Mpc}^{-3}}$, they constitute 10 to 20 per cent
of all clusters and groups with an X-ray luminosity greater than $2.5 \times
10^{42} h \;{\rm ergs\; s}^{-1}$ \cite[][]{ Vikhlinin1999a, Romer2000a,
  Jones2003a}. The number density of fossil groups is comparable to that of
brightest cluster galaxies \cite[][]{Jones2003a}. Thus they may be of
considerable importance as the place of formation of a significant fraction of
all giant ellipticals.
  
So far different approaches to model and measure the number density of fossil
systems have been undertaken. \citet[][]{Milosavljevic2006a} adopted an
extended Press-Schechter approach to estimate (5-40)\% for the expected
fraction of fossil groups in a mass range of $10^{13}-10^{14} \hMsun$ and
decreasing to (1-3)\% for massive clusters.  Recently
\citet[][]{vandenBosch2007a} used 2dF data to measure a fossil fraction of 6.5
\% for group with masses $10^{13}-10^{14}\hMsun$ . Using N-body simulations 
\citet[][]{dOnghia2007a} estimated a fraction of 18\% and 
\citet[][]{Sales2007a} (8-10)\% for $M_{\rm group} > 10^{13}\hMsun $ based on
an analyses of the Millennium simulation \cite[][]{Springel2005a}.  A
compilation of measurements of the fossil fraction and a discussion of
differences due to selection effects is given in \citet[][]{Dariush2007a}.
Note, however, that some of these systems seem to be fossil clusters rather
than fossil groups \cite[][Zibetti et 2007, in preparation]{Gastaldello2007a,
  Cypriano2006a}, i.e. galaxy clusters with the typical magnitude-gap of 2
between the brightest and the second brightest cluster galaxy. It seems at
least that a significant fraction of groups is fossil and it is a strong
function of group mass \cite[][]{Milosavljevic2006a, Sales2007a,
  vandenBosch2007a} .
 
Most of the theoretical work on fossil groups focused on the predictions of
the statistics of the magnitude-gap in the luminosity function. The physical
processes that lead to the formation of a mass or magnitude-gap in these
systems are still poorly understood. Early work suggested that fossil groups
result from mergers of the largest galaxies within compact groups
\cite[][]{Barnes1989a} and are due to early formation time
\cite[][]{dOnghia2005a,Dariush2007a}.  However, it is not yet understood under
which conditions mergers are so efficient that they produce such an extreme
gap in magnitude. When do fossil groups typically form their magnitude-gap?
Are fossil systems isolated systems or can they also populate high density
regions like galaxy clusters? Are fossil systems early formed systems, are
they more concentrated than other systems? Are fossil groups long lasting
systems, or does the group environment regulate its lifetime by infall of new
massive structures?
 
These are the open questions which we address here using a high-resolution
N-body simulation. The answers should guide the interpretation of
observational datasets especially by surveys like PANSTARS combined with
COSMOz that can search for fossil systems at higher redshift and provide a
framework for understanding the formation of giant ellipticals within the
current cosmology.
 
This paper is organized as follows. We describe the numerical simulation in
$\S2.1$ and the selection criterion of the sample of fossil groups in $\S2.2$.
  
In $\S3$ we describe the fossil group properties we find, like number density,
formation time, concentration and time of last major merger. We investigate
the formation mechanisms leading to the large magnitude-gap in $\S4$. Our main
results are summarized in $\S5$.

\section{Methods}

We have selected our sample of fossil groups from a 80 $h^{-1}{\rm Mpc}$ dark
matter only N-body simulation which is large enough to lead to a statistically
meaningful sample. Since we are mainly interested in the dynamical properties
of the massive group members, we focus on a dark matter simulation in this
work.

\subsection{Simulation}

The initial conditions were generated for a WMAP3 cosmology with matter
density $\Omega_{\rm m} = 0.24$, an linear mass variance on 8
$h^{-1}$Mpc-scale $\sigma_{8} = 0.76$, a dimensionless Hubble parameter $h =
0.73$ and a spectral index of primordial density perturbations $n = 0.96$
\cite[][]{Spergel2007a}. We used $N =512^{3}$ dark matter particles, i.e. a
particle mass of $4 \times 10^{8} h^{-1} {\rm M}_{\odot}$. Starting at
redshift $z = 40$ we evolved the simulation until the presence with the MPI
version of the Adaptive Refinement Tree (ART) code \cite[][]{Kravtsov1997a}.
The ART code enabled us to reach a force-resolution of $~1-3$ kpc in the most
refined regions, making sure our massive sub-haloes do not suffer from
over-merging \cite[][]{Klypin1999a} inside of our group sized haloes.

We identified groups with a friend-of-friends (FOF) algorithm with a linking
length of $l = 0.17\; d$ (corresponding to a mean over-density of roughly
330), with $d$ the mean inter-particle distance. The advantage is that it
identifies groups of any shape. In a second step we identified the bound
(sub-)structures in the groups.  To this end we used the Bound Density Maximum
(BDM) halo finder \cite[][]{Klypin1999a}. This algorithm removes unbound
particles from the haloes and is therefore particularly suitable to identify
sub-haloes and their properties, like their circular velocity. Since the
determination of sub-halo masses in groups is uncertain we characterized them
by their maximum circular velocity. The BDM halo with the highest circular
velocity within the FOF group is considered the host group halo. We found 116
groups in the mass range of $(1\times 10^{13} - 5 \times 10^{13}) h^{-1}{\rm
  M}_{\odot}$, corresponding
to the massive end of galaxy groups. 

We calculated the mass accretion histories of the haloes using 130 time steps
of equal distance in the expansion parameter, $\Delta a = 0.006$. To this end
we selected the 20 per cent of the most bound particles of haloes and compare
them in 8 consecutive time steps. We uniquely associated a halo to its
progenitor by considering halo pairs, which have the largest number of
particles in common and do not differ by more than a factor 5 in mass. The
last criterion was included to avoid spurious misidentification by sub-haloes
with their host halo.

\subsection{Selection of Fossil Groups}

Fossil groups are defined on the basis of a measured magnitude-gap between the
brightest and the second brightest group member. In our simulation we can only
identify the dark matter haloes which host the group galaxies. Currently
methods are being developed in order to relate galaxy luminosities to dark
matter haloes in a statistical way \cite[e.~g.~][]{Yang2004a, Vale2004a,
  Cooray2005a, Conroy2006a}. Here we follow this idea and associate
luminosities to the (sub-)haloes in the group. We assume simply that the most
luminous galaxy is the central galaxy of the group host halo with circular
velocity ${\rm v_{\rm circ,1}}$.  Consequently, the halo with the second
highest circular velocity ${\rm v_{\rm circ,2}}$ will host the second
brightest group galaxy.  Here the circular velocity ${\rm v_{circ}}$ is
always taken at the maximum of the rotation curve.
  
To model the magnitude-gap we adopt a similar approach as
\citet[][]{Milosavljevic2006a} where we relate the the halo circular velocity
to the luminosity of the central galaxy using an empirically measured mean
R-band mass-to-light ratio \cite[][]{Cooray2005a}. Assuming a Sheth-Thormen
\cite[][]{Sheth1999a} mass distribution function for the dark matter haloes
and a functional form as in Equation 1, that expresses the halo mass in
luminosity for the central galaxies, \citet[][]{Cooray2005a} fit the measured
R-band luminosity function of \citet[][]{Seljak2005a}. We convert our circular
velocities to luminosities by the relation

\begin{equation}
L\left(M\right) = L_{0} \left(\frac{M}{M_{0}} \right)^{a} \left[ b + \left( \frac{M}{M_{0}}\right)^{cd} \right]^{-1/d}
\end{equation}

\noindent with $L_{0} = 5.7\times10^{9} L_{\odot}$, $M_{0} = 2\times10^{11}
{\rm M_{\odot}}$, $a = 4$, $b = 0.57$, $c = 3.78$, $d = 0.23$, where we
substitute circular velocities for masses using the relation found by
\citet[][]{Bullock2001a}: $M/\left(h^{-1}\;{\rm M_{\odot}}\right) =
10^{\alpha}\cdot \left[{\rm v_{circ}}\right/\left({\rm km
    \;s^{-1}}\right)]^{\beta}$, with $\alpha = 4.3$ and $\beta = 3.4$.  We
then define fossil groups as having masses in the range of $(1\times 10^{13} -
5 \times 10^{13}) h^{-1}{\rm M}_{\odot}$ and a magnitude-gap $\mgap \ge 2$ mag
in the R-band .

Mass accretion onto haloes stops at the time when they become sub-halos of a
more massive object like a group. After infall they start to lose matter due
to tidal interactions. Since baryons tend to lie deeper in the potential well,
they will be less prone to get tidally stripped. Therefore, the total
luminosity is more likely to be related to the mass at infall \cite[see
e.~g.~][]{Kravtsov2004a}. Following this idea we characterize the sub-haloes
of the groups by their masses and circular velocities {\it at infall times}
onto the group. The choice of relating luminosities to the circular velocities
of halos at infall time has been motivated by recent successes in matching the
data by modeling the two- and three-point correlation functions
\cite[][]{Conroy2006a, Berrier2006a, Marin2007a}.

\section{Distribution and properties of fossil groups}

\subsection{Abundance}

In this section our main aim is to characterize the properties of the fossil
groups of our group sample in order to guide the interpretation of future
observational constraints. We begin by computing the abundance of fossil
systems in our catalog.

Assuming a magnitude-gap of $\mgap \ge 2$ (see dashed line in Figure 1) 24 per
cent of the groups of our catalog are classified as fossil,  corresponding to
a number density of $5.5 \times 10^{-5} h^{3}{\rm Mpc}^{-3}$.  This rate is
higher than previous estimates based on $N$-body simulations
\cite[][]{dOnghia2007a}, semi-analytic models \cite[][]{Sales2007a,
  Dariush2007a}, and observational estimates \cite[][]{ Vikhlinin1999a,
  Romer2000a, Jones2003a, vandenBosch2007a}, which all get a fraction of
around 10 per cent for groups in the mass range considered here. 
However, only 15 well studied fossil groups are known at present with X-ray
data. Therefore these  abundances present large uncertainties and  might be
well underestimated.  We believe that the over-estimate comes from our adopted
scheme for relating circular velocities to luminosities of the central
galaxies in groups, where we followed \citet[][]{Milosavljevic2006a} and
\citet[][]{Bullock2001a}.  We are interested mainly in the formation process
of systems with a large magnitude-gap, which clearly corresponds to systems
with a large gap in circular velocities even if the related magnitudes are
uncertain. We therefore stick to our adopted method and study how our selected
fossil population differs from the normal group population.

\begin{figure}
  \hspace{-1.5cm} \epsfig{file=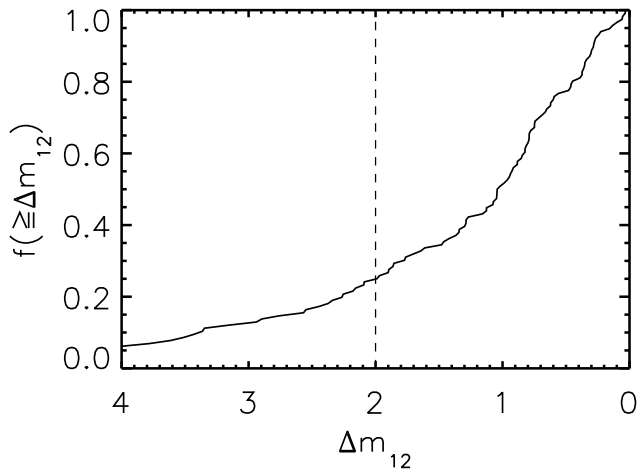,width=10cm}
 \caption{The fraction of galaxy-group sized haloes with a magnitude-gap
   parameter larger than $\mgap$. The dashed line indicates our defining
   magnitude-gap for fossil groups $\mgap \ge 2$.}
\label{vratio}
\end{figure}

\subsection{Environment Density Dependence}

Fossil groups are systems with many properties typical for galaxy clusters.
Hence, a further interesting test concerns the question whether fossil groups
are isolated systems that populate the low density regions or tend to reside
in higher density regions of the Universe like galaxy clusters. A good test
would be the cross-correlate the X-ray emitting fossil groups with galaxies in
the nearby universe, e.g with SDSS data. However the limited number of fossil
groups actually known makes an estimate of such correlations extremely
difficult. Some observational indications, though still uncertain, would
suggest that fossil groups could be fairly isolated systems
\cite[e.g.][]{Jones2000a, Jones2003a, Adami2007a}.

\begin{figure}
  
  \hspace{-1cm}\epsfig{file=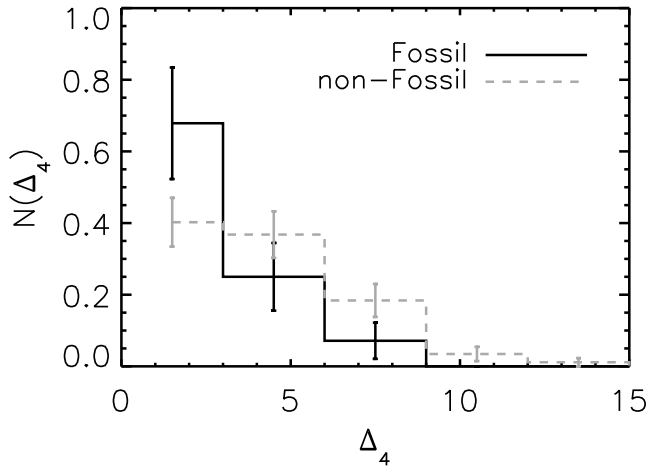,width=10cm}
 \caption{The group environment over-density $\Delta_{4}$ computed in a sphere of $ 4
   \hMpc$ for fossil (black solid line) and non-fossil groups (grey dashed
   line). To compute the over-density, the inner virial radius was
   subtracted. For comparison, cluster typically populate regions of
   $\Delta_{4} > 10$. There does not seem to be a strong tendency for fossil
   groups to be in low density environments. Error bars indicate 1$\sigma$
   Poissonian uncertainties.}
\label{density}
\end{figure}

We check in our simulated sample of groups whether fossil systems populate
preferentially low density regions in the universe. We estimate the
environmental density on a scale of 4 \hMpc. To this end we determine the
environmental over-density $\Delta_{4} = \rho_{4}/\rho_{\rm bg} - 1$, where
$\rho_4$ is the dark matter density within 4 \hMpc from the group center of
mass, with the inner one virial radius is subtracted, and $\rho_{\rm bg}$ is
the background matter density. Figure ~\ref{density} shows the distribution of
the over-density $\Delta_{4}$ for fossil and normal groups.  Most of the
groups, in the range of mass considered here, independent of being fossil or
not, populate preferentially the intermediate over-density region. A two-sided
Kolmogorov-Smirnov test for the two cumulative distributions shown gives
$D=0.31$, corresponding to a probability of 0.03 that the the two samples are
drawn from the same distribution.  We checked that adopting a slightly larger
or smaller volume (scales of 2-5 \hMpc) does not change our results
significantly. We do not find a strong tendency for fossil systems to be
preferentially located in low density environments.  We suggest therefore that
observations might be biased to find fossil groups preferentially in low
density regions, which could be due to group selection effects.

\subsection{Formation Time}

\begin{figure}
  \hspace{-1cm}\epsfig{file=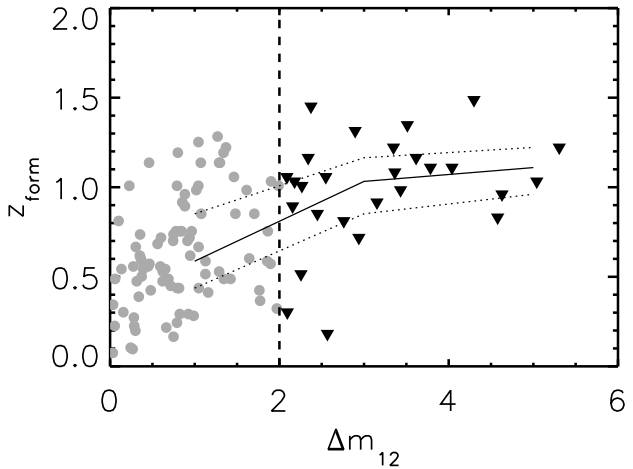,width=10cm}
 \caption{ The correlation between the formation redshift of the
   group host halo and its magnitude-gap parameter for fossil groups
   (triangles) and normal groups (circles). Over-plotted are the mean
 values (solid line)and lower and upper quartiles (dotted lines).} 
\label{lastmm}
\end{figure}

\begin{figure}
  \hspace{-1cm} \epsfig{file=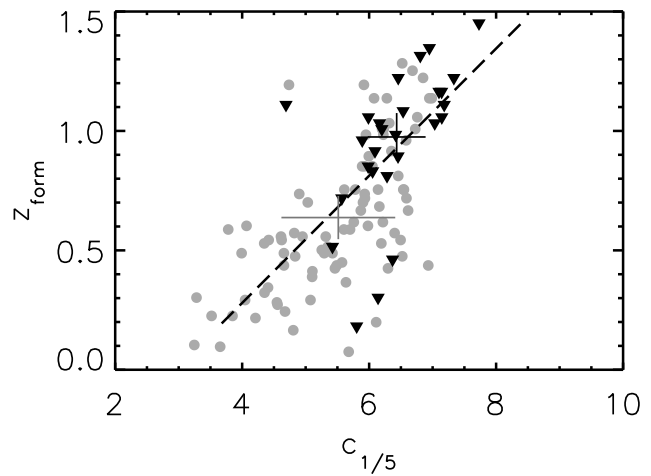,width=10cm}
 \caption{ 
   The formation redshift of the host halo as a function of the group
   concentration. The fossil group sample is marked with triangles and the
   normal groups are drawn with circles. The dashed line is a linear fit to
   all points. Crosses correspond to the mean values for the concentrations of
   fossil: $c_{1/5} = 6.4$ (bold cross) and non-fossil groups $c_{1/5} =5.5$
   (light cross), with the widths indicating the 1$\sigma$ standard deviations.}
\label{lastmm}
\end{figure}

The halo formation redshift is defined as the epoch at which the system
assembled 50\% of its final mass \cite[e.g.][]{Lacey1993a}. Figure 2 shows a
correlation between the magnitude-gap parameter and the formation redshift of
the host halo for all the fossil systems (triangles) and the normal groups
(grey circles). As already pointed out in \citet[][]{dOnghia2005a,
  dOnghia2007a} this correlation shows that fossil groups tend to form earlier
than normal groups, albeit with large scatter. In order to assess this
correlation we draw the mean and upper and lower quartiles (the solid and
dotted lines in Figure 2). The visual impression is quantified by statistical
measures as the Pearson's linear correlation coefficient $r = 0.39$, implying
a weak linear correlation between the magnitude-gap and the formation time.

\begin{figure*}
\begin{center}
  \hspace{-2cm}
\begin{minipage}[t]{1.0\textwidth}
 \begin{minipage}[t]{0.5\textwidth} 
   \epsfig{file=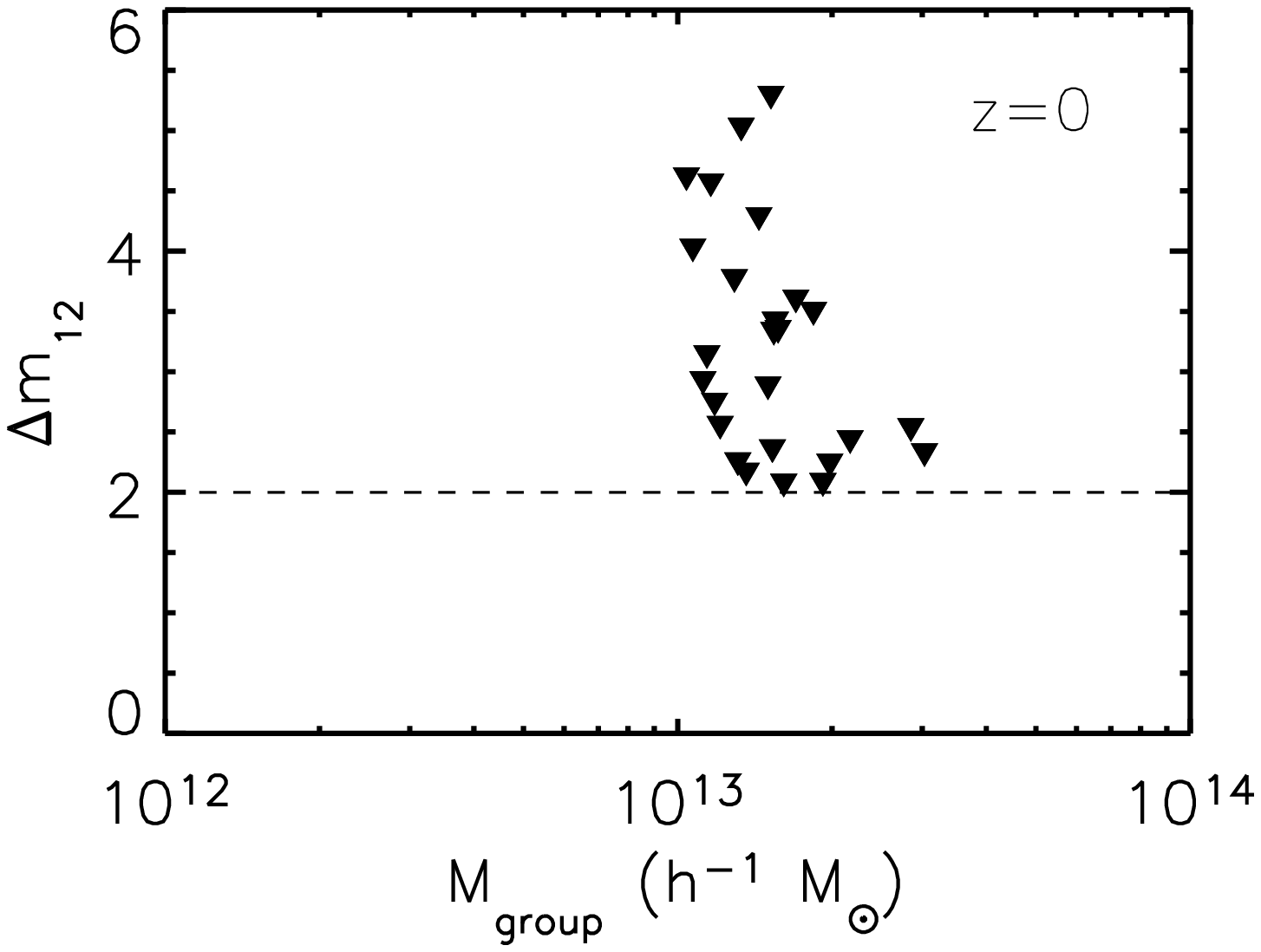,width=10cm}
 \end{minipage}
 \hfill
 \begin{minipage}[t]{0.5\textwidth} 
   \epsfig{file=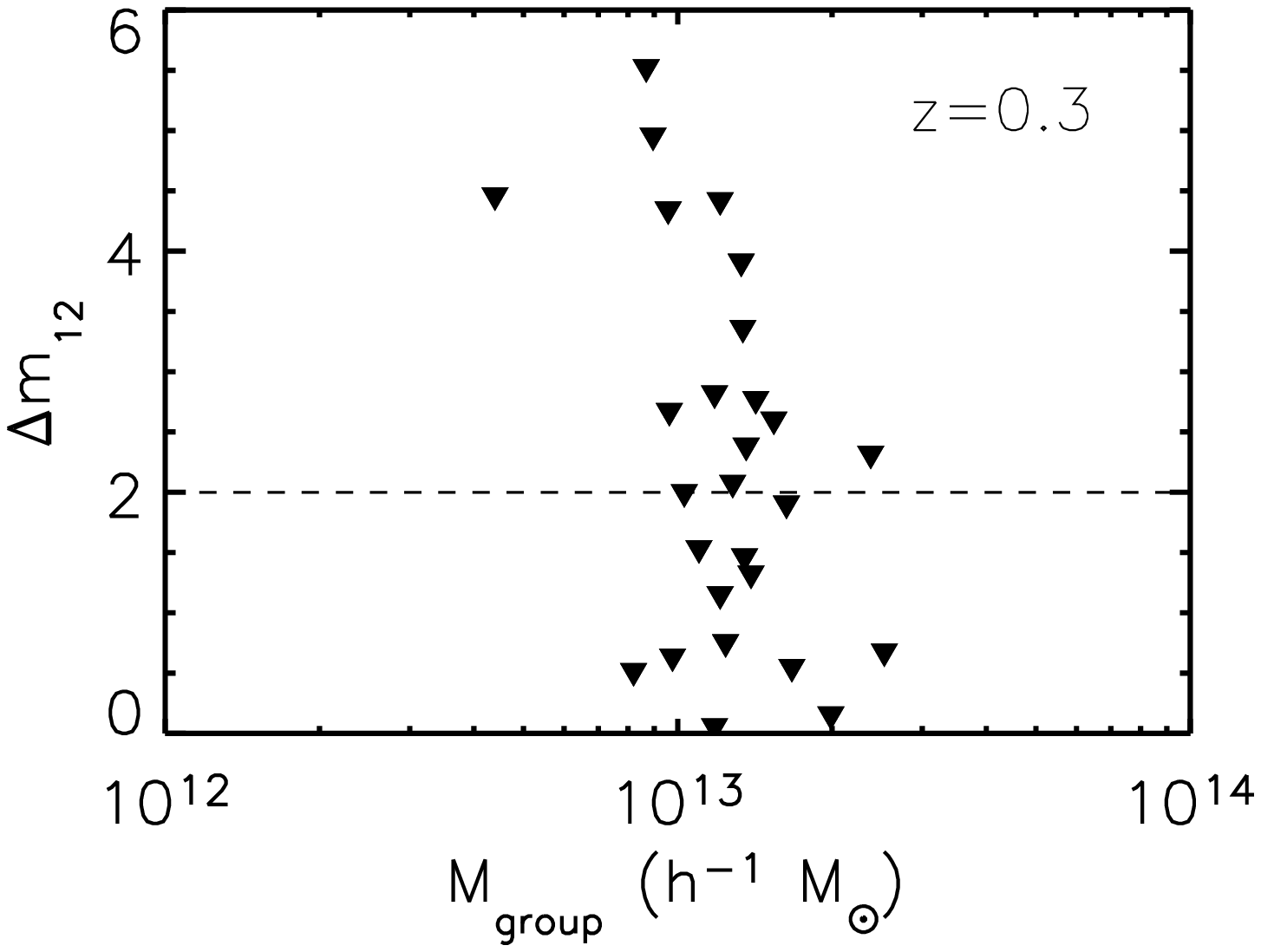,width=10cm}
 \end{minipage}
\end{minipage}

\vfill \hspace{-2cm}
\begin{minipage}[t]{1.0\textwidth}
 \begin{minipage}[t]{0.5\textwidth}
   \epsfig{file=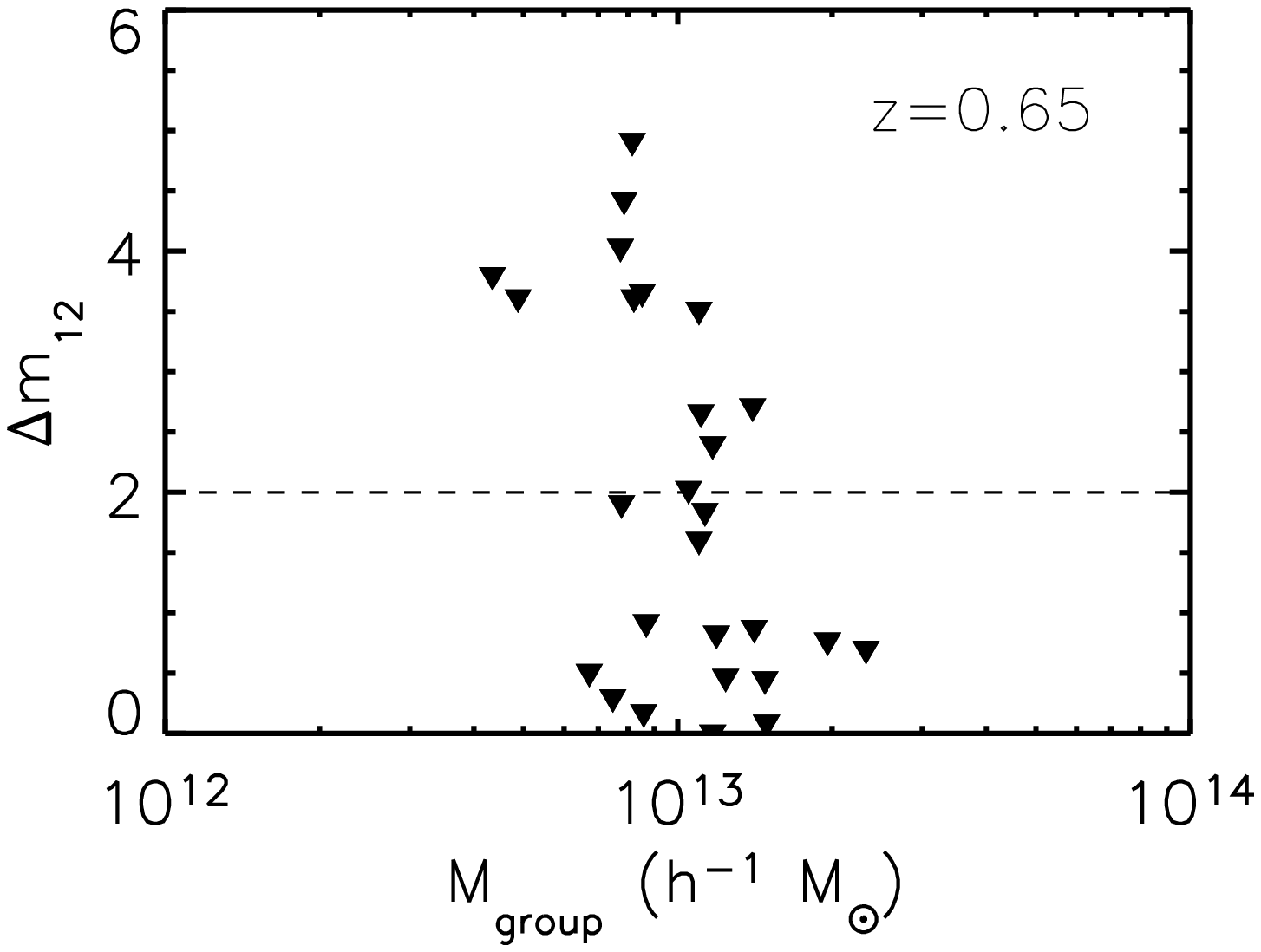,width=10cm}
 \end{minipage}
 \hfill
 \begin{minipage}[t]{0.5\textwidth} 
   \epsfig{file=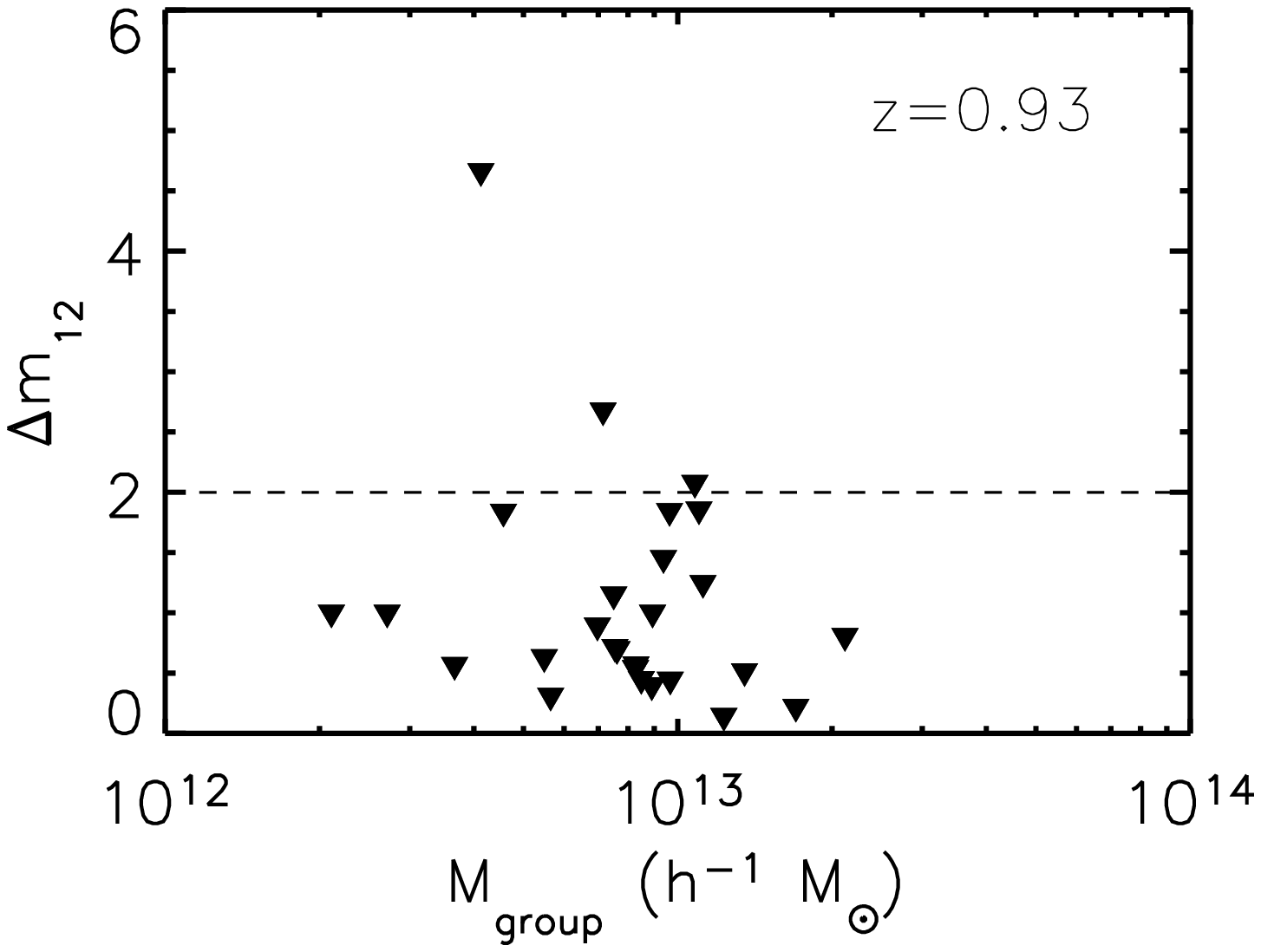,width=10cm}
 \end{minipage}
\end{minipage} 
\end{center}
 \caption{The evolution of the magnitude-gap parameter $\mgap$ for 
   fossil groups selected at $z$=0 (for redshifts $z$=0, 0.3, 0.65, 
     and 0.93). The non-fossil groups, which constitute the majority of
   the groups, are left out from this plot for clarity. The plot shows that the
   majority of the fossil systems had one or more massive satellites at higher
   redshift.  Note that the formation of the magnitude-gap happens typically
   later (between $z=0-0.7$) than the formation of the groups, which occurred
   around $z\ge0.8$ (see Figure 2).}

\label{vratiotime}

\end{figure*}
\begin{figure*}
\begin{center}
  \hspace{-2cm}
\begin{minipage}[t]{1.0\textwidth}
 \begin{minipage}[t]{0.5\textwidth} 
   \epsfig{file=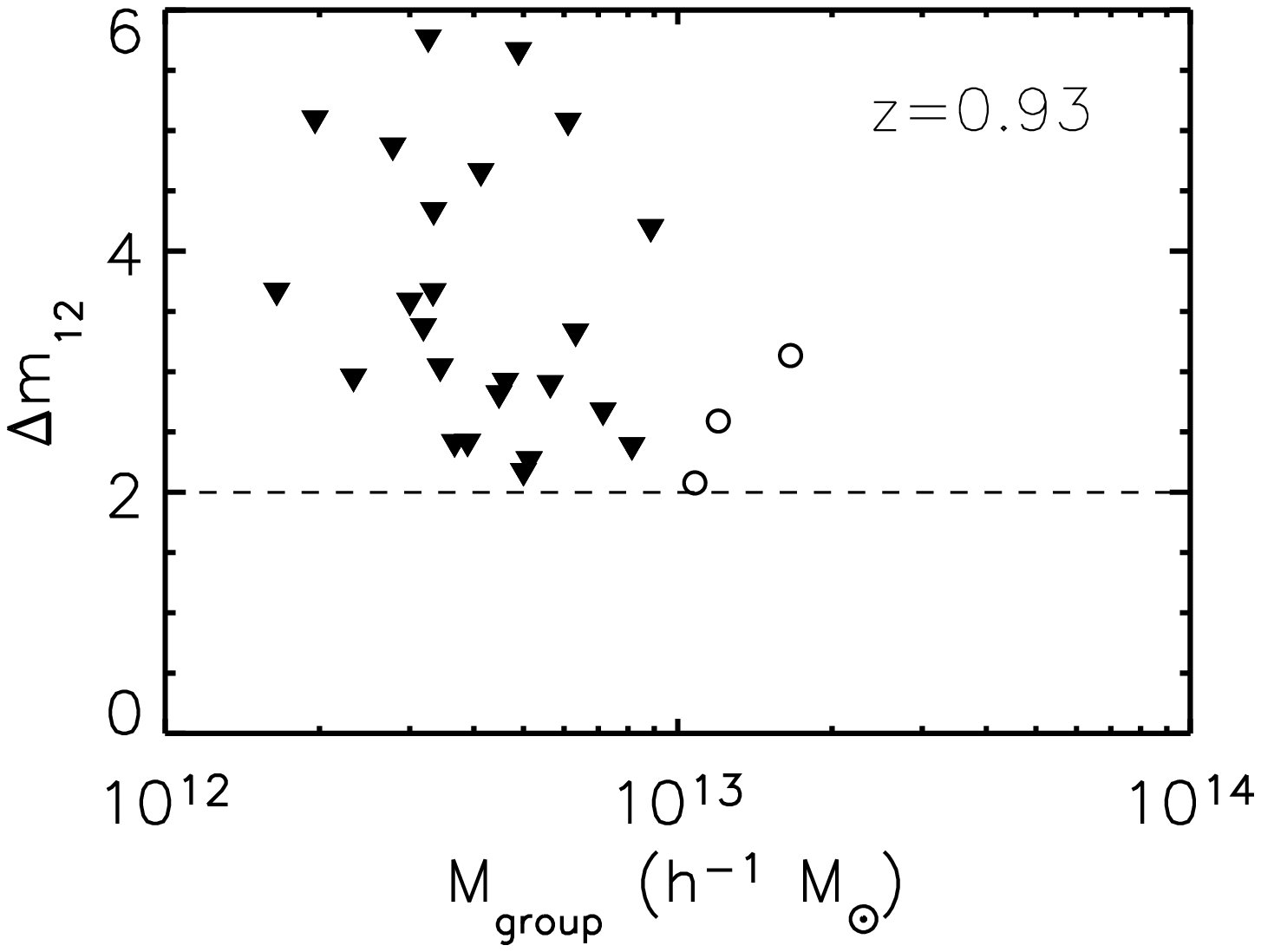,width=10cm}
 \end{minipage}
 \hfill
 \begin{minipage}[t]{0.5\textwidth} 
   \epsfig{file=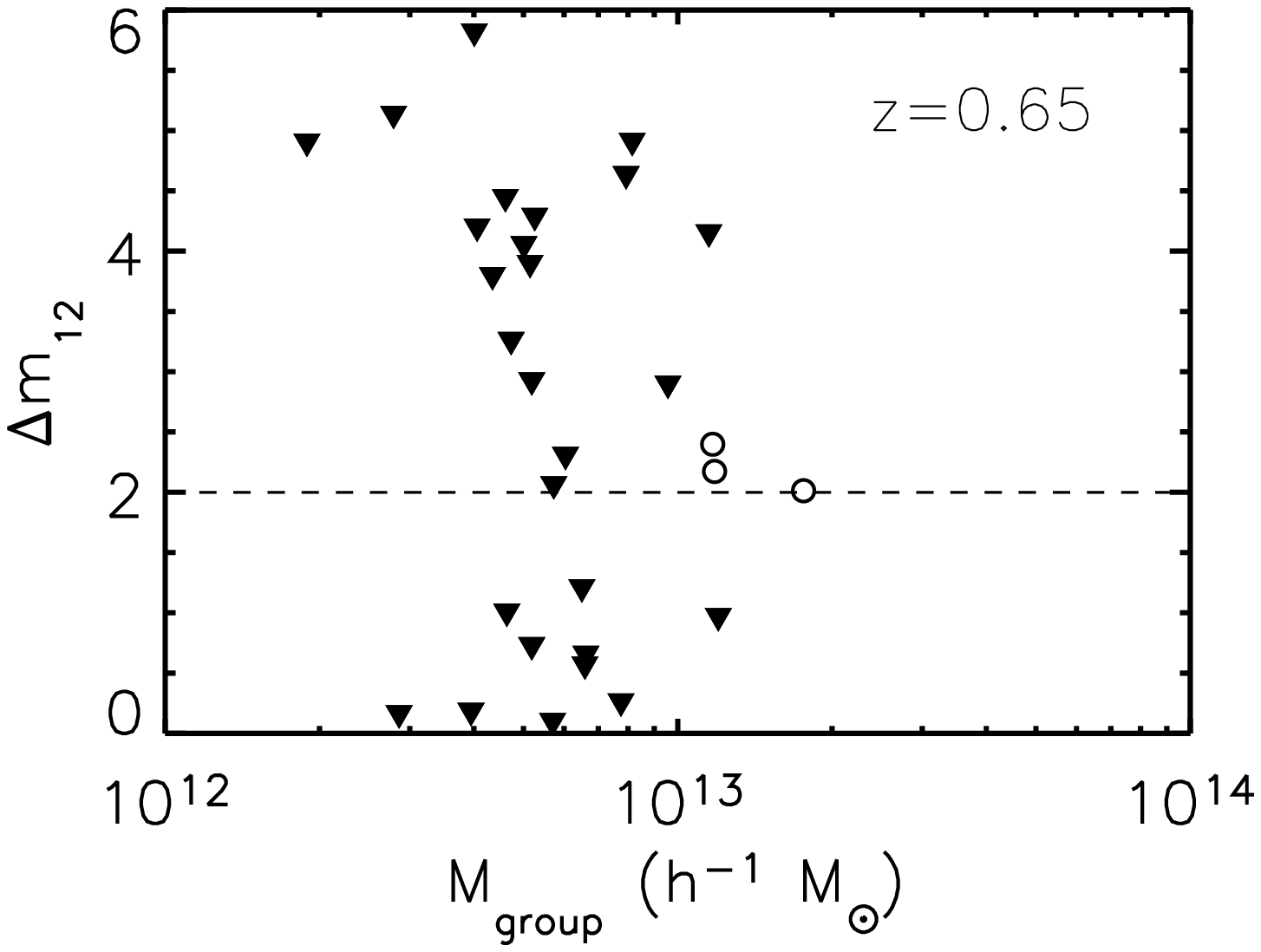,width=10cm}
 \end{minipage}
\end{minipage}

\vfill \hspace{-2cm}
\begin{minipage}[t]{1.0\textwidth}
 \begin{minipage}[t]{0.5\textwidth}
   \epsfig{file=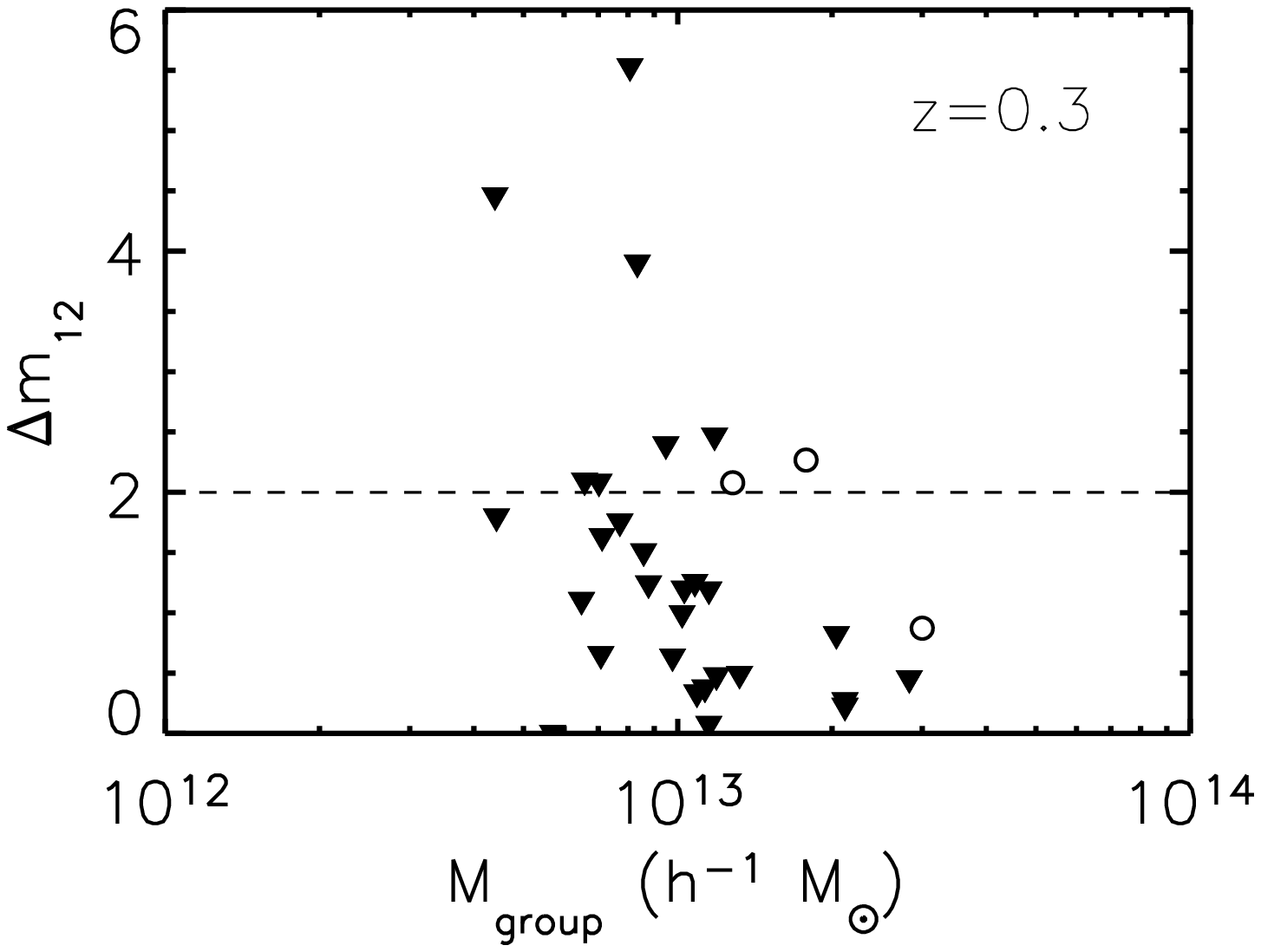,width=10cm}
 \end{minipage}
 \hfill
 \begin{minipage}[t]{0.5\textwidth} 
   \epsfig{file=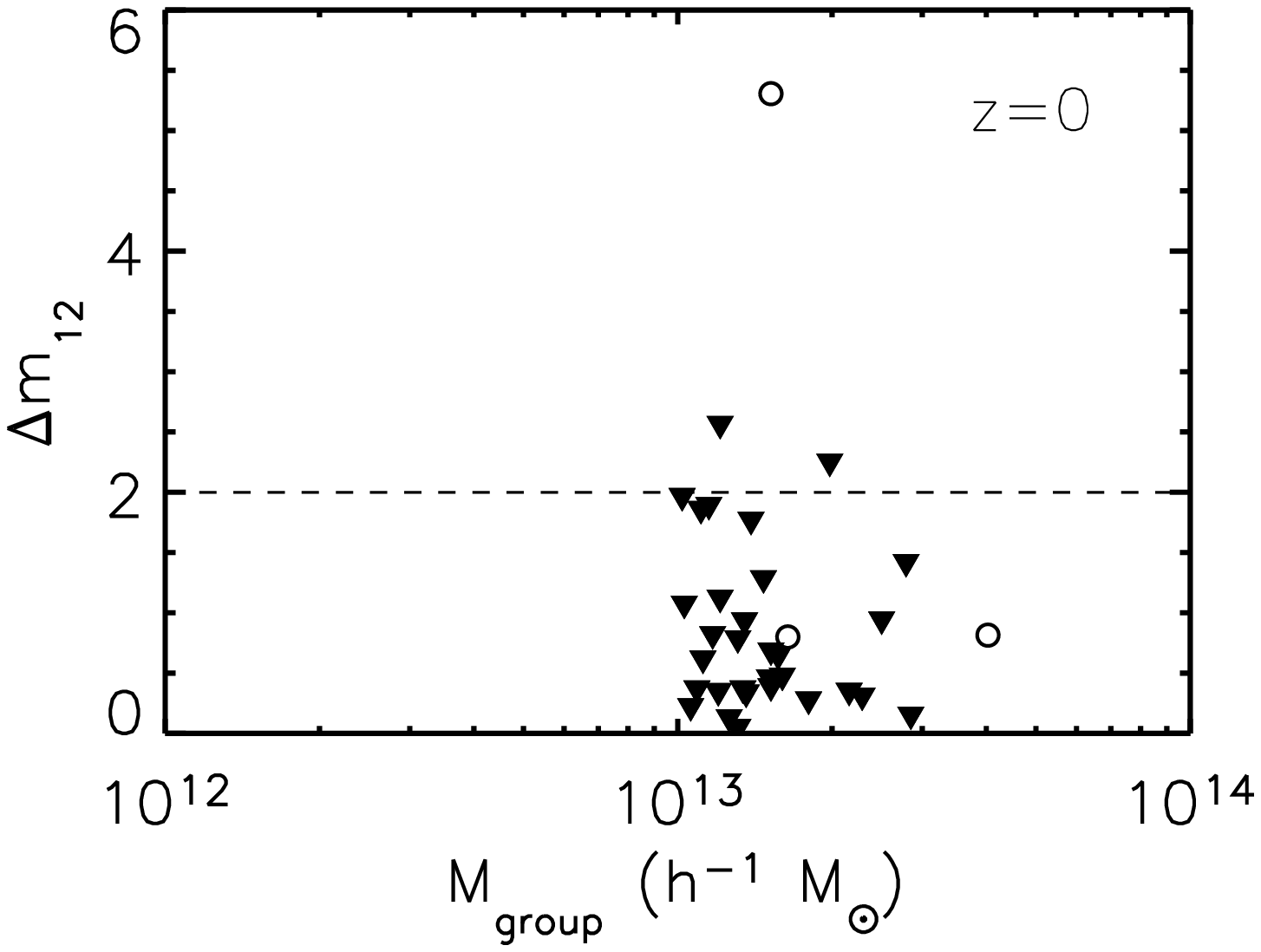,width=10cm}
 \end{minipage}
\end{minipage} 
\end{center}
 \caption{The evolution of the magnitude-gap parameter $\mgap$ for 
   fossil groups selected at $z$=0.93 that end up in the $z$=0 group sample,
   given at different epochs ($z$=0.93, 0.65, 0.3, and 0). Again
   for clarity, only the selected sample at $z$=0.93 is shown. The open
   circles indicate fossil groups that are selected in the same mass range as
   the group sample at $z$=0. Most of the high redshift fossil systems, when
   traced forward in time, experience renewed infall of massive satellites and
   become normal groups at $z$=0. These systems are undergoing a `fossil
   phase'. }

\label{vratiotime}
\end{figure*}

\subsubsection{Concentration Parameter}

The early formation redshift is also reflected in a higher concentration
parameter \cite[][]{Navarro1997a}.

We define the concentration of our haloes by the ratio of the virial radius of
the host halo to the radius of a sphere enclosing one fifth of its virial
mass: $c_{1/5} = r_{\rm vir}/r_{1/5}$. This definition of the halo
concentration allows for a robust concentration determination when the haloes
are merger remnants and un-relaxed \cite[][]{AvilaReese2005a}. The correlation
shown in Figure 3 between formation redshift and concentration is well fitted
by a linear relation $z_{\rm form} = -0.79 + 0.27 \times c_{1/5}$ (marked with
the dashed bold line).

The fossil groups clearly populate the early formed, more concentrated part of
the plot, and have a mean concentration of $c_{1/5} = 6.4$, which is about 16
per cent higher than the concentrations found for normal groups $c_{1/5} =
5.5$. Our findings are consistent with fossil groups being systems with higher
dark matter concentrations than usual groups, which supports such a trend
found by \citet[][]{Khosroshahi2007a}. At present there are yet few
observational constraints on this issue \cite[e.g.][]{Gastaldello2007a,
 Khosroshahi2007a}. However with the upcoming X-ray observations of fossil
groups with Chandra and XMM will provide soon better constraints.

\subsection{Last Major Merger}

Recent studies of the giant elliptical at the center of fossil groups report
no signs of a recent major merger activity, indicating that any major merger
should have happened at least more than approximately 3 Gyrs ago
\cite[][]{Jones2000a, Khosroshahi2006a}.

For each halo we estimate the time of the last major merger of the group
haloes of our sample by studying the detailed mass assembly history. To identify
the time of the last major merger, we denote a halo as a major merger remnant
if its major progenitors were classified as a single group at one time but two
separate groups with a mass ratio less than 4:1 at the preceeding time. Note
that when the mass ratio defining the major merger event is restricted to
almost 1:1 progenitors, the time of the last major merger should in general
coincide with the formation time (as defined above). We find that only 15\% of
the fossil groups experienced the last major merger less than 2 Gyrs ago, and
at least 50\% had the last major merger longer than 6 Gyrs in qualitative
agreement with the observations.

\section{Formation of Fossil Groups}

In previous sections we investigated some properties of our galaxy-group sized
haloes that can be compared to observations. In particular, the fossil groups
appear to be more concentrated systems, they formed earlier, with the last
major merger happened a long time ago, in qualitative agreement with current
observational constraints. We now turn to the question why these systems are
devoid of a significant substructure. Obviously, no large satellite has fallen
in lately.  The question then is, how long is such a major infall ago -- if it
ever occurred? And if they fell in, when do they merge to create the magnitude-gap?

Furthermore, we look for signs of efficient merging in fossil systems. We
focus on two main conditions that lead to the formation of these systems: i)
the low infall rate of massive satellites into the host halo; and ii)
distribution of angular momentum of the in-falling massive satellites.

\subsection{Forming the Magnitude-Gap}

To quantify when the magnitude-gap was formed in fossil systems, we plot the
distribution of the magnitude-gap of fossil groups selected at $z=0$ in Figure
5. This sample is traced backward in time to $z=0.3$ (the right top panel);
$z=0.65$ (the bottom panel on the left) and $z=0.93$ (the bottom panel on the
right). We find that the groups selected as fossil at present, show a lower
magnitude-gap once traced backwards in time. They therefore do not qualify
anymore as fossil systems at higher redshifts. The magnitude-gap is typically
formed over a wide range of redshifts between $z=0-0.7$. It worth noting that
this happens typically after the group halo has gained half of its mass, which
occurred earlier around redshift $z \ge 0.8$ (see Fig.~ 2).

\subsection{The Fossil Phase}

Is the `fossil stage' a final stage or will the systems fill their
magnitude-gap with the time? To assess this question, we select a sample of
fossil systems at high redshift $z=0.93$, and track these forward in time (as
shown in Figure 6). The open circles indicate massive fossil groups in the
same mass regime as the sample selected at $z=0$. Following all the fossil
systems forward in time we note that they leave the range where they would be
identified as fossil systems due to new in-falling satellites. Only three of
these systems did not experience further infall of a massive satellites from
their surrounding environment so that they end up as a fossil system today.

Our simulation suggest that the existence of a gap in the galaxy luminosity
function in fossil systems is only a transition phase in the evolution of
groups, the duration of which is related to the merging of group members with
the central object and infall of new haloes.

\subsection{Properties of In-falling Satellites}

\begin{figure}
  \hspace{-1cm}\epsfig{file=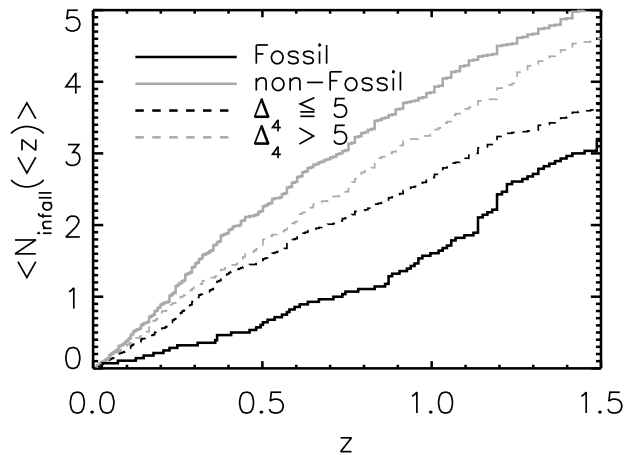,width=10cm}
 \caption{The mean cumulative number of massive satellites with 
   $\mgap < 2$ mag in-falling into the host halo at redshifts $z_{\rm infall}
   < z$. The grey solid line corresponds to normal groups and the black solid
   line to fossil groups. The dashed curves indicate the mean cumulative
   number of massive satellites when the group samples are split up by low
   dense regions $\Delta_{4} \le 5$ and higher dense regions $\Delta_{4} > 5$.
 }
\label{n_infall}
\end{figure}

We showed in previous sections that fossil groups tend to assemble earlier
their mass than normal groups, and that the unusual magnitude-gap is a
transient phase in the evolution of the group in the hierarchical universe. A
group identified as fossil today did not experience any late infall of massive
satellites to fill the gap in the magnitude distribution function of the group
members. In order to assess this question quantitatively we compute the
average infall rate of haloes onto fossil systems as compared to normal
groups.  This quantity is expressed in our analysis by computing the
cumulative number of massive (within 2 magnitudes of the host) satellites
falling into the host halo after redshift z divided by the total number of
fossil (normal) groups: $<N_{\rm infall}\left(<z\right)> = N_{\rm
  infall}(<z)/N_{\rm groups}$.  Figure~\ref{n_infall} shows that fossil groups
accrete the larger satellites earlier in time as compared to the normal
groups. An early infall of massive companions ensures enough time for the
massive satellite to merge into the host halo. Note that for $z>0.8$ fossil
and non-fossil groups have similar infall history of massive satellites. In
fact the slopes of the evolution of the infall rate is the same for both
distribution for $z>0.8$, showing that the infall rate of satellites is only
different at low redshifts.

We checked whether the difference in infall rate is determined by environment.
This is done by splitting the sample up in $\Delta_{4} < 5$ and $\Delta_{4}
\ge 5$ and evaluating the cumulative number of in-falling satellites (see
dashed lines in figure 7). Although the denser regions do experience a bit
more infall, the difference is only of the order of 10-30 per cent, not as big
as we observe for fossil and non-fossil systems (in agreement with
\citet[][]{Maulbetsch2007a}, who find little environmental dependence of the
mass accretion history in this mass range). This supports lack of a strong
environmental preference for fossil groups found in section 3.2.

\begin{figure}
  \hspace{-1cm}\epsfig{file=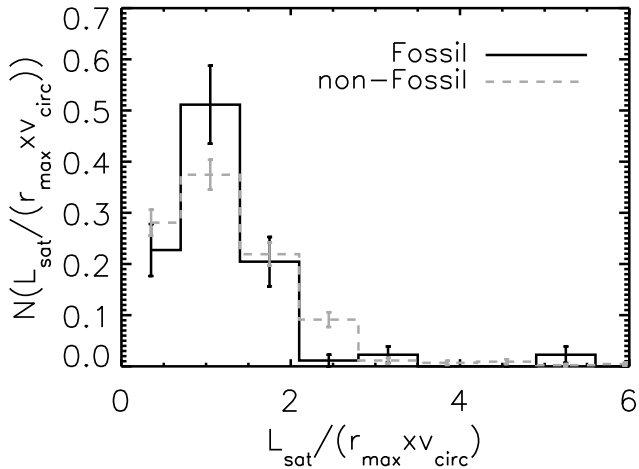,width=10cm}
 \caption{ The distribution of the in-falling satellite normalized angular
   momentum for fossil (black solid line) and non-fossil groups (grey dashed
   line). The distribution seems more narrowly distributed around $L_{\rm sat}
   \approx r_{\rm max}{\rm v_{\rm circ}}$ for fossil groups, i.e. its high
   angular momentum tail seems to be less pronounced. Error bars indicate
   $1\sigma$ Poissonian uncertainties.}
\label{angmom}
\end{figure}

We address the question whether the conditions under which the massive
satellites enter the group enable an efficient merger which could lead to a
gap in the magnitude distribution of the fossil group galaxies. We checked the
angular momenta values of the satellites at time of infall.
Figure \ref{angmom} shows the angular momentum of the satellites at infall
time. The angular momentum is calculated by taking the cross product of the
distance at infall and the velocity at infall $L_{\rm sat} = \left({\bf r_{\rm
      inf}}\times{\bf v_{\rm inf}}\right)/\left(r_{\rm max}\cdot{\rm v_{\rm
      circ}}\right)$, with $r_{\rm max}$ and ${\rm v_{circ}}$ both measured at
the maximum of the rotation curve. Both distributions peak at $L_{\rm sat}
\approx r_{\rm max}{\rm v_{\rm circ}}$. Fossil groups seems to be lacking
satellites in the high end tail of angular momentum distribution, which may
cause a faster merging of the satellites.  However, since the distribution is
rather poorly sampled, better number statistics are needed to confirm this
result.

\section {SUMMARY AND CONCLUSIONS}

From an N-body simulation, we select and analyze a large sample of galaxy
group-sized haloes, that allows us to study in detail the mechanisms that lead
to the formation of fossil systems in the hierarchical universe.

Our criterion to select fossil systems is based on identifying galaxy-group
sized haloes showing a gap in the magnitude between the two most massive
members. This selection criterion assumes that the circular velocity of a halo
traces its luminosity until it becomes a substructure of another more massive
system. We relate the magnitude of our haloes to the circular velocities by
using the empirical mean relation between dark matter halo mass and central
galaxy R-band luminosity found by \citet[][]{Cooray2005a}.

Our results may be summarized as follows:

\begin{itemize}
\item In the mass range $ 1\times 10^{13} - 5\times 10^{13} h^{-1}{\rm
    M}_{\odot}$ 28 of the 116 groups sized haloes, i.~e.~ 24 per cent qualify
  as fossil groups according to the definition of a magnitude-gap of 2
  magnitudes between the brightest and second most bright galaxy.  This
  fraction is higher than the measured values.  The largest uncertainty here
  is how to relate the mass or circular velocity of the haloes to the
  luminosities of their central galaxies. Because our adopted method of
  relating mass to galaxy luminosity is uncertain due to the broad scatter in
  this relation \cite[][]{Cooray2006a}, and the rate is sensitive to the group
  definition, as well as selection effects, a robust comparison to observed
  number densities obtained by other authors is difficult at the moment.  We
  are selecting systems with large circular velocity- or respectively
  mass-gaps, so that our sample can be used to study the formation of the
  extreme magnitude-gaps observed in fossil systems.
 
\item The fossil groups identified in our sample tend to form earlier than the
  other groups. The fossil systems have assembled half of their final mass
  around $z\ge 0.8$ in agreement with the previous works. They form their
  magnitude-gap typically between redshift $z=0-0.7$, much later than the
  formation time of the groups. The early formation time is also expressed in
  a slightly higher dark matter concentration. The average concentration for
  the fossil groups is $c=6.4$ compared to $c=5.5$ of normal groups. Further,
  we find that the majority of the fossil group seem to have experienced the
  last major merger longer than 3 Gyr ago.
 
\item We do not find a strong correlation between the environment and the
  formation of fossil systems.  Observations that indicate that fossil systems
  are found preferentially in low density environments might be biased by
  selection effects.
 
\item The primary driver for the large magnitude-gap is the early infall of
  massive satellites that is related to the early formation time of these
  systems. The difference in infall rate for different group environments is
  only of order 10-30 per cent, far less than the observed difference between
  fossil and non-fossil groups, and hence the current environment does not
  seem the primary driver for the lack of massive satellites. This is in
  agreement also with the lack of strong correlation between fossil systems
  and environment.
 
\item We suggest that efficient mergers of massive members within the groups
  can create the magnitude-gap typical of fossil groups at any redshift. The
  efficiency of the merger process seems to be linked to the lower angular
  momentum of the massive satellites in-falling into the host halo of fossil
  groups when compared to normal groups. However, due to the limited number
  statistics we need more data to substantiate this.
  
\item By selecting samples of fossil groups at higher redshift ($z \approx 1$)
  we find that many of them do not exhibit the magnitude-gap anymore once they
  are traced forwards until present time. The majority of them fill the
  magnitude-gap with time by infall of new massive satellites. We conclude
  that the stage for a group to be ``fossil'' is a transient phase.
\end{itemize}

\section*{Acknowledgments}

The computer simulation was done at Columbia supercomputer at NASA Ames
Research Center.  A.v.B.B. acknowledges support from Deutsche
Forschungsgemeinschaft (DFG) under the project MU 1020/6-3. E.D. is supported
by a EU Marie Curie Intra-European Fellowship under contract MEIF-041569. A.K.
acknowledges support of NSF grant AST-04070702. M.H.  acknowledges support
from the DFG under the project Vo 855/2.

\bibliography{articles_foss} \bsp

\label{lastpage}

\end{document}